# Two dissociable axes of brain hierarchical dynamics track universal and population-specific effects of mindfulness.

Victor Oswald1,2, David Zarka3, Riku Johannes Ihalainen1,2, Gulshan Kumar1,2, Rikkert Hindriks4, Auretta Sonia Kummar5, Helen Correia5,6, Hakuei Fujiyama5, Saketh Malipeddi7, Ravindra P Nagendra7, Arun Sasidharan7, Bindu M Kutty7, Steven Laureys*1,2, Prejaas K.B. Tewarie*1,2

1. Joint International Research Unit on Neuroplasticity, CERVO Brain Research Centre and Centre de recherche du CHU de Québec, Université Laval, Quebec City, QC, Canada

2. CERVO Research Institute, Laval University, Québec, Canada

3. Unité de Recherche en Sciences de l'Ostéopathie, Faculté des Sciences de la Motricité Humaine, Université Libre de Bruxelles

4. Department of Mathematics, Vrije Universiteit Amsterdam, The Netherlands

5. School of Psychology, College of Health & Education, Murdoch University, Perth, Western Australia, Australia

6. Psychological Sciences, Australian College of Applied Professions, Perth, Western Australia, Australia

7. Centre for Consciousness Studies, Department of Neurophysiology, National Institute of Mental Health and Neurosciences, Bengaluru, India

* co-shared last authors

**Corresponding authors**:

Victor Oswald– CERVO Research Institute, Laval University, Québec, Canada 2525 Chemin de la Canardière, Québec, QC G1J 2G3, Canada

Email: victor.oswald@umontreal.ca

Ravindra P Nagendra- Centre for Consciousness Studies, Department of Neurophysiology, National Institute of Mental Health and Neurosciences, Bengaluru, India

Email: ravindrapinna@gmail.com

**Author Contributions:** V.O. conceptualized the study, designed the analysis framework, implemented the sliding-window TII pipeline, the trajectory-descriptor extraction, the joint Hidden Markov Model, and the behavioural bridge analyses; performed all statistical analyses; produced all figures; and wrote the original draft. P.K.B.T. and S.L. co-supervised the project and provided conceptual and methodological guidance throughout; contributed to the interpretation of dynamical-systems and clinical findings respectively; and edited the manuscript. R.H. developed the original Temporal Irreversibility Index framework, provided conceptual input on its interpretation, and contributed to methodological validation. D.Z. provided the MBSR cohort (data acquisition, preprocessing, ethical clearance) and contributed to methodological input on longitudinal EEG analysis in mindfulness research. A.S.K., H.C., and H.F. provided the CoMET trauma-exposed cohort (study design, data acquisition, clinical assessment, ethical clearance) and contributed to the interpretation of the trauma-related findings. S.M., R.P.N., and B.M.K. provided the ISHA Yoga cohort (recruitment at the Isha Yoga Center, data acquisition, cohort characterization, psychometric assessment, ethical clearance) and contributed to the interpretation of the meditation-expertise findings. R.J.I. and G.K. contributed to methodological development of the TII pipeline, source-reconstruction and Brainstorm homogenization, and to code review and reproducibility validation. All authors reviewed and approved the final manuscript.

**Competing Interest Statement:** The authors declare no competing interests.

**Classification:** Biological Sciences, Neuroscience. Dual classification: Social Sciences, Psychological and Cognitive Sciences.

**Keywords:** Temporal Irreversibility Index; brain dynamics; mindfulness; Mindfulness-Based Stress Reduction; Hidden Markov Model; Post-Traumatic Stress Symptoms; behavioural neuroplasticity

**This PDF file includes:**

Main Text
Figures 1 to 5
Supplementary Information (Supp Fig. S1 and Supp Fig. S2)

**Abstract**

How mental training reshapes the temporal architecture of large-scale brain dynamics and how these mechanisms differ between healthy and clinically vulnerable populations remains poorly understood. Time-averaged neural measures collapse the multi-scale nature of brain activity, obscuring the dynamic processes through which behavioural interventions reorganise hierarchical brain function. Here, applying a sliding-window Temporal Irreversibility Index to source-reconstructed EEG across three complementary cohorts (n = 141), we decompose hierarchical brain trajectories into three orthogonal temporal descriptors and uncover two dissociable, mechanistically interpretable axes of intervention-induced plasticity. Cumulative meditation practice drives a universal, pan-cerebral reduction in the exploratory breadth of hierarchical dynamics, tracking dose-dependently with lifetime hours of practice. Temporal persistence is bidirectionally tuned by 8-week mindfulness-based interventions in a population-specific way: reduced in healthy meditators but increased in trauma-exposed individuals, revealing that comparable MBI protocols engage qualitatively distinct network mechanisms depending on baseline clinical status. A joint Hidden Markov Model further links these dynamical shifts to distinct microstate substrates, and longitudinal changes selectively track improvements in a single facet of mindfulness, attentional presence in daily life. Together, these findings establish a mechanistic framework for how behavioural training reorganises non-equilibrium brain dynamics, with direct implications for tailoring mindfulness-based interventions to individual clinical profiles.

**Significance Statement**

Mindfulness-based interventions are among the most widely deployed non-pharmacological treatments for anxiety, depression, and trauma-related disorders, yet how they reshape brain dynamics and how those mechanisms differ between healthy and clinical populations has remained unclear. Applying a temporal decomposition of hierarchical brain dynamics to EEG from three complementary cohorts (n = 141), we identify two dissociable axes of intervention-induced neuroplasticity: a universal dose-dependent narrowing of exploratory breadth with cumulative practice, and a bidirectional tuning of temporal persistence that reverses direction between healthy meditators and trauma-exposed individuals. These findings establish population-specific neuroplasticity as a general principle for tailoring behavioural interventions across clinical profiles.

## Main Text

### Introduction

Behavioural interventions such as mindfulness-based training have well-documented effects on attention, emotion regulation, and clinical symptoms (1–4). Their neural correlates have been extensively mapped using time-averaged measures, fMRI amplitudes, static functional connectivity, and EEG power spectra, revealing modulation of default-mode, salience, and attentional networks (5–8). Yet the brain operates simultaneously across many timescales, from millisecond-scale sensory processing to second-scale cognitive transitions. Capturing how interventions reshape this multi-scale temporal organization requires moving beyond static averages.

The stakes of this question extend well beyond basic neuroscience. Mindfulness-based interventions are now among the fastest-growing non-pharmacological treatments for anxiety, depression, chronic pain, and trauma-related disorders, and are being integrated into occupational, educational, and public-health programs worldwide (1, 3). Yet the mechanisms through which mental training exerts these behavioural effects remain poorly characterised at the neural level, limiting our ability to (i) predict who will benefit, (ii) match specific interventions to individual clinical profiles, and (iii) design next-generation contemplative programs with mechanistic targets. Progress here requires moving from descriptive neural correlates to a mechanistic understanding of how behavioural training reshapes the temporal architecture of large-scale brain function.

A growing literature reframes cognition as the temporal evolution of large-scale neural dynamics (9–12). One theoretically grounded probe of this temporal structure is the Temporal Irreversibility Index (TII), which measures the asymmetry of neural cross-correlations under time-reversal (13). In physical systems, time irreversibility has been argued to reflect departure from thermodynamic equilibrium; in the brain, it indexes directional information flow and hierarchical asymmetry between top-down and bottom-up interactions (14, 44). While static TII discriminates levels of consciousness (15), neurodegenerative disease (16), and pharmacological states (17), it collapses the temporal dimension and quantifies the average hierarchical depth, but conceals how that depth unfolds over time: whether the brain lingers in specific hierarchical modes, switches rapidly between them, or fluctuates broadly.

Here, we address the neuroscientific gap concerning how mental training reorganises multi-scale brain dynamics by leveraging TII as a temporal probe. By computing TII within a sliding window, we obtain a continuous trajectory of hierarchical states that captures how the brain unfolds across time rather than a single time-averaged summary. From each trajectory, we extract three orthogonal descriptors: the mean depth of hierarchical asymmetry (indexing the average level of top-down versus bottom-up flow), the exploratory breadth of hierarchical states (indexing how widely the brain samples its dynamic repertoire), and the temporal persistence of the current configuration (indexing how smoothly it evolves over time). Together, these three parameters characterise a dynamic regime, analogous to how statistical physics characterises stochastic processes through mean, variance, and correlation time.

Mindfulness-based interventions (MBIs) provide a uniquely informative window into how behavioural training reshapes brain dynamics, because they place naive brains at one end of a well-characterised training gradient and long-term expert practitioners at the other. Although prior work has argued that meditation stabilizes brain dynamics (18), reduces default-mode variability (19), or promotes critical regimes (20), these accounts rely on single cohorts, single timescales, or metrics that are difficult to compare across studies. Three core questions therefore remain open: (i) does mindfulness-based training modify the temporal architecture of hierarchical brain dynamics beyond its average level? (ii) are these changes dose-dependent with cumulative practice? and (iii) what state-level mechanisms underlie these dynamical shifts?

To answer these questions, we combined three independent EEG cohorts (n=141 total): a cross-sectional Isha Yoga cohort across three expertise levels (advanced, novice, and non-meditators), a longitudinal 8-week Mindfulness-Based Stress Reduction (MBSR) cohort in healthy adults, and a clinical cohort of trauma-exposed individuals with Post-Traumatic Stress Symptoms (PTSS) undergoing Compassion-oriented and Mindfulness-based Exposure Therapy (CoMET) (22). Across all cohorts, we probed six cognitively relevant timescales spanning 40 to 600 ms.

We demonstrate that mental training reshapes brain hierarchical dynamics along two dissociable axes: a universal, dose-dependent reduction in exploratory breadth, and a population-specific modulation of temporal persistence that diverges between healthy and trauma-exposed individuals. A joint Hidden Markov Model demonstrates that the exploratory-breadth axis is a state-occupancy phenomenon, whereas hierarchical depth and temporal persistence index complementary temporal features of the state dynamics not summarised by fractional occupancy. Crucially, longitudinal changes in both dynamical markers specifically track gains in mindfulness acting-with-awareness. Together, these findings establish the temporal architecture of hierarchical brain dynamics as a mechanistically decomposable and behaviourally anchored substrate of mental training.

## Results

To probe how mindfulness-based interventions reshape brain dynamics across timescales, we computed the sliding-window Temporal Irreversibility Index (TII) on source-reconstructed EEG from three complementary cohorts (n = 141): a cross-sectional Isha Yoga cohort (n = 96, three expertise levels: advanced meditators, novices, and non-meditator controls), a longitudinal 8-week Mindfulness-Based Stress Reduction (MBSR) cohort in healthy adults (n = 28), and a longitudinal 8-week Compassion-oriented and Mindfulness-based Exposure Therapy (CoMET) cohort in trauma-exposed individuals with Post-Traumatic Stress Symptoms (PTSS; n = 17). For each subject and each of six probed timescales ($M \in \{40, 72, 152, 252, 352, 600\}$ ms), we obtained a trajectory of the 28 inter-network TII couplings across the seven canonical Yeo functional networks (Schaefer-100 parcellation). We then summarised each trajectory by three orthogonal descriptors: its temporal mean (average hierarchical depth), temporal variance (exploratory breadth), and lag-1 autocorrelation (temporal persistence). All statistics used BH-FDR correction with correction unit specified per analysis (see Methods).

Hierarchical dynamics reveal dissociable axes of trait stabilization and intervention response

To evaluate whether meditation training targets distinct facets of hierarchical dynamics, we contrasted the three trajectory descriptors, hierarchical depth, exploratory breadth, and temporal persistence , across two independent experimental paradigms (Fig. 1a-c): a cross-sectional expertise cohort of meditation (Isha Yoga: advanced practitioners [ADV, median 4,850 h lifetime practice], novices [NOV, 1,450 h], and controls [CNT, 0 h]; n=96) and two longitudinal within-subject interventions (healthy MBSR [n=28] and CoMET (trauma-exposed) [n=17], pre- vs. post-intervention). Average hierarchical depth (hierarchical depth; Fig. 2, top row) showed no significant differences across expertise levels in the cross-sectional cohort (Welch's t-test $q>0.29$) and remained unchanged following either longitudinal intervention (paired sign-flip permutation, $q>0.15$ for both MBSR and CoMET). This triple null finding highlights the stability of baseline hierarchical depth (visualised across all three descriptors and six lags in Supplementary Fig. 2) and underscores the selective sensitivity of the remaining two dynamic metrics. Exploratory breadth (exploratory breadth; Fig. 2, middle row) displayed a monotonic gradient across expertise levels (ADV < NOV ≤CNT) across five of six probed timescales, four of which survived false discovery rate correction (BH-FDR; 152 ms: $t=-2.86, q=0.018$; 252 ms: $q=0.018$; 352 ms: $q=0.018$; 600 ms: $q=0.022$). Longitudinally, exploratory breadth decreased consistently across all six lags in healthy meditators undergoing MBSR (paired-permutation $P<0.09$ across lags). Conversely, the trauma-exposed cohort undergoing CoMET exhibited an opposing trend (positive t-values across all lags, non-significant after FDR correction), establishing the specificity of exploratory breadth reduction to standard mindfulness practice. These findings suggest that mindfulness training constrains the exploratory repertoire of hierarchical configurations over time, representing a trait-like axis of stabilization. Temporal persistence ( Fig. 2, bottom row) did not differ across cross-sectional expertise groups ($q>0.90$) but exhibited robust, bidirectional shifts following longitudinal interventions. In healthy meditators, temporal persistence decreased significantly across all six lags (paired-permutation $q\leq 0.038$; peak at 72 ms: $t=-3.96, q=0.002$). In contrast, trauma-exposed individuals with PTSS showed significant increases in temporal persistence across three lags (40 ms: $q=0.019$; 252 ms: $q=0.026$; 352 ms: $q=0.026$). The opposing trajectories of these two clinical profiles suggest convergence toward a balanced dynamical regime, indicative of a bidirectional attractor axis. Notably, this axis is unobservable in cross-sectional expert practitioners, who have already stabilized at long-term equilibrium. Together, these results establish a three-part decomposition of hierarchical dynamics: an invariant baseline axis (hierarchical depth), a trait-like stabilization axis (exploratory breadth), and an intervention-driven attractor axis whose directional modulation depends on clinical baseline (temporal persistence ).

Mindfulness alters network dynamics globally in healthy individuals but focally in trauma

To determine whether changes in exploratory breadth and temporal persistence reflect global network reorganization or region-specific targeting, we evaluated couplings across 28 Yeo-7 network pairs at each timescale (Fig. 3). Cross-sectional effects were quantified using Cohen's d (ADV vs. CNT), while longitudinal changes were assessed via paired effect sizes (d_z). Reductions in exploratory breadth were pan-cerebral in healthy cohorts. At the 152 ms reference lag, all 28 network pairs exhibited significant reductions in advanced practitioners compared to controls ($q<0.05$). This distributed pattern was maintained across adjacent timescales (72 ms and 252 ms). Healthy participants post-MBSR demonstrated a concordant pan-cerebral reduction, strongest at 39–72 ms (26 of 28 pairs with uncorrected $P<0.05$). Conversely, trauma-exposed individuals with PTSS showed elevated exploratory breadth tendencies across most network pairs, with focal increases reaching trend significance in limbic–sensorimotor circuits (Fig. 3c). Modulations of temporal persistence were similarly pan-cerebral in healthy meditators (26 of 28 network pairs showing FDR-significant reductions post-MBSR) but highly localized in trauma-exposed individuals. In the trauma-exposed cohort, temporal persistence increases were restricted to dorsal attention couplings with sensorimotor, executive control, and default-mode networks, alongside intra-limbic connections. This spatial divergence indicates that MBSR engages distinct network mechanisms depending on baseline health: inducing widespread neural flexibilization in healthy individuals while driving targeted renormalization of hypervigilance and limbic circuits in PTSD.

HMM state-level analysis: distinct dynamical substrates for the three axes and a population-specific dissociation

To elucidate the microstate architecture underlying these macroscopic trajectory descriptors, we trained a joint 5-state Hidden Markov Model (HMM) on the TII trajectories at the primary timescale (M=152" ms" ) across all cohorts (n=141unique subjects, 186 subject × condition segments; Fig. 4a). Model selection across $K\in\{3,\dots,7\}$ identified $K=5$ as optimal on the basis of free-energy minimization and Hamming-distance consistency across random initializations (Supplementary Fig. 2). The five inferred states exhibit highly persistent dynamics (diagonal self-transition probabilities 0.67"--" 0.84; Fig. 4a1–a2), separated by sparse inter-state transitions. To prevent non-independence introduced by paired pre/post samples in the longitudinal cohorts, we evaluated the association between per-subject state Fractional Occupancy (FO) and the three trajectory descriptors exclusively within the cross-sectional ISHA cohort (n=96, one observation per subject), applying Benjamini–Hochberg FDR correction across all 90 tests (5 states × 6 lags × 3 descriptors). Only exploratory breadth is captured by state occupancy, and with high magnitude: FO of State 3 correlates positively with exploratory breadth across four consecutive coarse lags (152"--" 600" ms" ; $\rho\approx+0.60$, all $P<0.001$), whereas FOs of State 2 and State 4 contribute equally strong negative associations over the same lags ($\rho\approx-0.60$, all $P<0.001$), with State 1 exhibiting a weaker negative trend (ρ up to -0.42 at 600" ms" , P<0.01). Subjects who spend more time in the sparse, transient integrative State 3 and less time in the persistent "anchor" States 2 and 4 display systematically higher trajectory variance, establishing exploratory breadth as a state-occupancy phenomenon. By contrast, hierarchical depth and temporal persistence remain confined within a narrow band around zero across all 5 states and 6 lags (with only State 4 at the fastest lag of 40" ms" reaching significance: P<0.01 for hierarchical depth, P<0.05 for temporal persistence ). Thus, hierarchical depth and temporal persistence are not captured by static FO; instead, they are encoded in complementary temporal properties of state-space dynamics (e.g., within-state dwell-time distributions, transition rhythms, or phase relationships). Paired sign-flip permutation testing on pre/post FO within each longitudinal MBSR cohort (, n=28; CoMET, n=17), corrected for multiple comparisons across the 10 pre/post contrasts (5 states × 2 cohorts, BH-FDR), revealed opposing state-level reorganizations within the two low-occupancy states. Following the 8-week MBSR protocol, healthy MBSR meditators exhibited a decrease in State 1 occupancy alongside an increase in State 2 occupancy (q<0.05). Conversely, trauma-exposed patients (CoMET) displayed the inverse pattern: an increase in State 1 and a decrease in State 2 (uncorrected P<0.05, non-significant after FDR at n=17). States 3, 4, and 5 showed no significant intervention-related shifts in either cohort. Critically, the integrative State 3, the primary FO substrate of trait exploratory breadth identified in the cross-sectional cohort, was unmodulated by either 8-week intervention. This confirms that the exploratory breadth trait effect observed in ISHA reflects long-term consolidation of state occupancy rather than short-term, practice-induced state shifts. Instead, acute intervention effects concentrate on the two low-occupancy states (State 1 and State 2) that reorganize in opposite directions depending on baseline clinical status. This state-level divergence directly underpins the population-specific temporal persistence response (Fig. 2b; reduction in MBSR vs. increase in CoMET) as well as the contrast between pan-cerebral and focal spatial patterns (Fig. 3).

Hierarchical dynamic shifts systematically track cumulative practice and mindfulness facets

Finally, we evaluated whether these neural trajectory shifts correlate with behavioral metrics and practice history (Fig. 5). In the cross-sectional cohort (n=96), log-transformed lifetime practice hours (log_10[1+"hours" ]) correlated negatively with exploratory breadth across four consecutive timescales (72 ms: ρ=-0.30,P<0.05; 152 ms: ρ=-0.33,P<0.01; 252 ms: ρ=-0.28,P<0.05; 352 ms: ρ=-0.25,P<0.05; Fig. 5a). This continuous dose-response relationship across orders of magnitude, from naive controls to practitioners with over 15,000 hours, provides strong observational support linking cumulative mental training to constrained hierarchical dynamics. In the pooled longitudinal MBSR sample (n=43), individual changes in brain dynamics (Δ exploratory breadth and Δ temporal persistence ) were correlated against changes across the five subscales of the Five-Facet Mindfulness Questionnaire (FFMQ). Longitudinal reductions in exploratory breadth correlated with overall FFMQ gains (peak at 600 ms: ρ=+0.43,P<0.05) and specifically with the Acting-with-Awareness subscale across five timescales (peak at 600 ms: ρ=+0.47,P<0.05; Fig. 5b). Changes in temporal persistence selectively tracked Acting-with-Awareness across all six timescales (peak at 352 ms: ρ=+0.48,P<0.01). No other FFMQ facet demonstrated consistent FDR-corrected associations. Scatter plots at peak timescales (Fig. 5c) confirm that participants undergoing larger neural shifts exhibited corresponding improvements in daily attentional awareness (Δ exploratory breadth at 600 ms vs. Δ Acting-with-Awareness : ρ=+0.47,P=0.003; Δ temporal persistence at 352 ms vs. Δ Acting-with-Awareness : ρ=+0.48,P=0.0015). Healthy and clinical participants aligned on a single regression slope, demonstrating that this behavioral anchor is invariant to baseline clinical status..

## Discussion

By tracking the temporal dynamics of source-reconstructed EEG across three complementary cohorts, we demonstrate that mindfulness-based training reshapes the multi-scale temporal architecture of hierarchical brain dynamics along two dissociable, behaviourally-relevant axes. Rather than acting as a monolithic modulator of neural activity, the implicit assumption of most single-metric contemplative-neuroscience studies to date (2–4, 6, 7), mental training selectively constrains the exploratory breadth of hierarchical states as a function of cumulative lifetime practice, while bidirectionally tuning temporal persistence according to baseline clinical status. A joint Hidden Markov Model further reveals that only the breadth axis is anchored in the fractional occupancy of discrete brain states, whereas the depth and persistence axes live in complementary temporal features not summarised by static state occupancy. Together, these findings reframe mindfulness-driven neuroplasticity as an orthogonal decomposition of hierarchical brain dynamics, an axis of trait-level consolidation, an axis of population-specific attractor convergence, and an invariant setpoint, and establish a mechanistic framework for how behavioural training reorganises non-equilibrium brain dynamics.

<u>Mindfulness-based interventions engage distinct network mechanisms in healthy versus trauma-exposed individuals.</u>

Spatial and state-level analyses converge on a striking population-dependent divergence in response to comparable 8-week MBI protocols. In healthy individuals in the MBSR cohort, mindfulness training drives a systemic, pan-cerebral flexibilization across all 28 network couplings, extending the local finding of decreased default-mode connectivity in experienced meditators (5, 19) to the full connectome. In trauma-exposed individuals (CoMET), by contrast, adaptations are highly localised, concentrated within dorsal-attention, executive-control, default-mode, and limbic circuits, the same networks that show abnormal baseline activity and connectivity in PTSD (30, 31) and that are engaged by top-down emotion regulation (27, 29). Critically, temporal persistence shifts in the opposite direction in the two populations: it decreases in healthy MBSR meditators but increases in trauma-exposed individuals with PTSS after CoMET.

This directional divergence is consistent with a growing convergence in the clinical literature. Whereas in healthy adults mindfulness typically down-regulates default-mode and salience-network activity (5, 8, 19, 29), in PTSD the default-mode network is often hyper-connected at rest and requires normalisation rather than further suppression (30–32). Analogously, top-down amygdala-prefrontal coupling, which strengthens with mindfulness in healthy adults (18, 25, 26), needs to replace rather than deepen the pathologically labile hypervigilant loops in trauma-exposed populations. Under this framework, mindfulness does not apply a uniform global shift; it may stabilise pathologically labile circuits in clinical populations while enhancing fluid exploration in healthy brains. Our temporal-persistence result provides a mechanistically interpretable, cohort-invariant read-out of this divergence, a common attractor axis reached from opposite baseline states. This has direct translational implications for adaptive trial design (29) and for matching contemplative interventions to the biotype of individual patients.

<u>Skill acquisition reconciles short-term expansion and long-term consolidation.</u>

Comparing cross-sectional expert data with longitudinal intervention outcomes reveals an apparent biphasic profile: cumulative lifetime practice predicts a narrower dynamic range (reduced baseline exploratory breadth), whereas short-term MBSR gains in mindfulness correspond to maintained or increased dynamic exploration ($\Delta$ exploratory breadth × $\Delta$ Acting-with-Awareness, $\rho = +0.47$, $p = 0.003$). This mirrors classical models of motor and cognitive skill learning, which describe an initial expansion of the functional repertoire during early acquisition followed by long-term structural pruning and consolidation into a refined, low-variance regime (33–35). Comparable expansion-then-consolidation dynamics have been documented for gray-matter volume in juggling and language learners (33, 34), for cortical excitability in motor training, and for EEG microstate repertoires in expert vs novice performers (36). Consistent with this consolidation view, expert meditators also show reduced long-range temporal correlations and altered 1/f neural noise as a signature of practice maturation (43), suggesting that consolidation of hierarchical dynamics extends beyond TII descriptors to broader temporal-scaling properties.

Under this framework, short-term training encourages flexible state exploration, whereas years of practice consolidate these configurations into a refined, low-variance baseline regime. This resolves an otherwise puzzling directional discrepancy in the meditation-EEG literature, where short-term studies have reported increases in dynamic variability (19, 20) and long-term expert studies have reported decreases (5, 6). Recognising this biphasic trajectory is clinically relevant: attempting to constrain dynamic variability prematurely in novice practitioners may hinder early learning, whereas supporting flexible exploration during initial training may foster superior long-term outcomes (2, 3).

Acting-with-Awareness provides a specific behavioural anchor for dynamic shifts.
Across the five facets of the Five-Facet Mindfulness Questionnaire (FFMQ; 37), longitudinal changes in brain dynamics correlated selectively with Acting-with-Awareness, the capacity to maintain attentional presence during daily activities, at all six probed timescales ($\rho \approx +0.47$–$0.48$, all $q < 0.01$). The reactive and descriptive dimensions (Observing, Non-Judging, Non-Reactivity, Describing) showed no robust association with either dynamic marker. This selectivity aligns with the sub-literature identifying Acting-with-Awareness as the FFMQ facet most consistently linked to executive-attention performance, metacognitive control, and reduced mind-wandering (2, 38, 39), and echoes the "focused-attention" phenotype in the Lutz et al. taxonomy of contemplative practices (2). Our findings establish a concrete temporal-dynamics signature corresponding specifically to this core attentional component of mindfulness training, providing a mechanistic bridge between the phenomenology of trait mindfulness (37) and the network-dynamics literature on attentional control (10, 11).

Clinical translational implications.
The mechanistic dissociation we report has three concrete implications for the clinical translation of mindfulness-based interventions. First, the pan-cerebral versus focal spatial signatures suggest that biomarker-driven treatment matching may be feasible: a distributed baseline signature could indicate readiness for standard MBSR-style training, while a focal fronto-limbic profile, as observed in trauma-exposed individuals, may benefit from trauma-adapted interventions that specifically target hypervigilant circuits (22). Second, the population-specific direction reversal on the temporal-persistence axis (reduced in healthy meditators but increased in trauma-exposed individuals converging toward a common attractor) argues against a one-size-fits-all logic: the same MBI can push brains toward a shared regime from opposite baselines, which reframes clinical outcome expectations and monitoring strategies. Third, because the exploratory-breadth axis scales continuously with lifetime practice over four orders of magnitude (0 to > 15,000 hours), it provides a first candidate neural dose-response metric for contemplative interventions, potentially useful as an objective correlate for triaging when a given dose is sufficient or when additional practice is likely to yield diminishing returns. Realising these translations will require prospective clinical trials with pre-registered biomarker prediction, integration with wearable EEG platforms that can extend beyond laboratory settings, and cross-modal validation against behavioural, physiological, and clinical outcomes. Collectively, our findings position the temporal architecture of hierarchical brain dynamics as a mechanistically decomposable, scalable, and clinically-actionable target for the next generation of contemplative interventions.
Limitations and future directions

Several considerations qualify these findings.
First, although we applied a fully unified computational pipeline across the three cohorts (see Methods), residual hardware-level differences, including EEG recording systems (ANT Neuro 128-channel for MBSR, HydroCel Geodesic 128-channel for CoMET, and a distinct clinical acquisition setup for ISHA) and geographical sampling contexts, preclude direct quantitative pooling of raw effect sizes across cohorts without prior data-harmonisation procedures (40). Second, while the inclusion of the active-control clinical trauma-exposed cohort (CoMET; 22) provided a compelling test of population-specific mechanisms, both longitudinal cohorts (MBSR and CoMET) lacked a passive waitlist control, so future work using such designs will help isolate non-specific practice effects from mindfulness-specific ones (3). Third, our joint HMM used a covariance-only emission model to isolate dynamic dispersion; incorporating mean-activity and spectral features in future models, as done by the OSL-dynamics group in HCP resting-state data (11, 12), may reveal complementary functional structures. Fourth, formally testing the biphasic learning model we propose will require multi-year longitudinal designs with dense intermediate sampling, a design challenge that no current contemplative-neuroscience dataset satisfies (3, 7). Fifth, the Isha Yoga practice and the MBSR/CoMET protocols share the core element of attentional-monitoring meditation but differ in structure, dose, and secondary components (mantra, breath regulation, compassion focus); the cross-cohort convergence on trait-level exploratory-breadth reduction suggests a shared mechanism, but claims of full equivalence across practices would require dedicated head-to-head trials. Sixth, our TII implementation follows Hindriks and colleagues (13, 14), but alternative irreversibility estimators (Cofré, de la Fuente; 41, 42) may provide complementary or partially overlapping views of hierarchical asymmetry.

### Conclusion

Mindfulness-based training reshapes the multi-scale temporal architecture of hierarchical brain dynamics along two dissociable, behaviourally-anchored axes: a universal, dose-dependent narrowing of exploratory breadth with cumulative practice, and a population-specific bidirectional tuning of temporal persistence. At the state level, only the breadth axis is a fractional-occupancy phenomenon; the two other axes index complementary temporal features of the state-space dynamics. Longitudinally, both dynamical shifts selectively anchor to gains in attentional presence during daily life, the acting-with-awareness facet of mindfulness. By revealing global network reorganisation in healthy meditators alongside targeted fronto-limbic stabilisation in trauma-exposed individuals, this work establishes the temporal architecture of hierarchical neural dynamics as a mechanistically decomposable and clinically-relevant substrate of mindfulness-based interventions, and offers a general framework for how behavioural training reshapes non-equilibrium brain dynamics across populations.

## Materials and Methods

### Participant cohorts and experimental design

Three independent resting-state EEG datasets were analysed under a unified computational pipeline (n = 141 total participants: 96 + 28 + 17).
Isha Yoga cohort (cross-sectional). Data from 96 participants recruited at the Isha Yoga Center (Coimbatore, India) as part of a larger study (n = 103). The sample comprised three expertise levels: advanced meditators (ADV, n = 38; median lifetime practice 4,850 h, range 953–15,400 h), novice practitioners (NOV, n = 30; median 1,450 h, range 250–4,255 h), and non-meditating controls (CNT, n = 28; 0 h). Each participant underwent four resting-state conditions centred around a standardised meditation session (PreMed, PostMed) and a control sit-still session (PreSitStill, PostSitStill), each containing eyes-closed and eyes-open sub-segments. Analyses focused on concatenated eyes-closed segments (≈ 8 min effective recording per subject, pooled across the four conditions). Healthy MBSR cohort (longitudinal). 28 healthy adults (10 female, 18 male; mean age 41.7 ± 10.9 years) completed a standard 8-week Mindfulness-Based Stress Reduction (MBSR) program (1). High-density resting-state EEG (128 channels, ANT Neuro) was acquired pre- and post-intervention (21). Trauma-exposed cohort (longitudinal clinical, CoMET). 17 trauma-exposed patients (mean age 27.0 ± 7.5 years) with clinically-significant Post-Traumatic Stress Symptoms (PTSS) completed an 8-week Compassion-oriented and Mindfulness-based Exposure Therapy (CoMET) program (22). High-density resting-state EEG (128 channels, HydroCel Geodesic Sensor Net) was recorded pre- and post-intervention. All original protocols received approval from their respective institutional ethics review boards, and informed consent was obtained from all participants prior to testing.

### EEG preprocessing and source reconstruction

All datasets were processed through a unified pipeline in Brainstorm (v2023). Raw signals were notch-filtered at 50 Hz, band-pass filtered between 1 and 60 Hz, and cleared of ocular and cardiac artefacts using Independent Component Analysis (Picard algorithm, 20 components). Cleaned sensor data were projected into source space using minimum-norm estimates (MNE) and parcellated according to the 100-region Schaefer cortical atlas with 7-network functional ordering (23). Regional time series were extracted by averaging source signals across vertices within each region of interest (ROI) and downsampled to 250 Hz. To prevent channel-reconstruction bias across recording sessions in the Isha Yoga dataset, channel templates were standardised using a single template-reference geometry across all subjects.

Sliding-window Temporal Irreversibility Index

Following Hindriks et al. (13), for each pair of ROI time series (x, y) we define the bivariate TII across maximum lag M as

$$\xi(x,y,M) = \frac{1}{M\sqrt{1-r_0^2}}\sum_{m=1}^{M}\left(r_m(x,y) - r_{-m}(x,y)\right),$$

where r_m (xⓜ,y)is the cross-correlation between x and y at lag m, and r_0is the zero-lag correlation. TII matrices (100 × 100 ROIs) were estimated within sliding time windows of duration 5M samples with a step of M/2 (90 % window overlap). ROI-level matrices were aggregated into 27-dimensional feature vectors representing pairwise couplings between the seven canonical Schaefer networks (23), yielding an n_"windows" ×28trajectory per participant × condition. Six maximum-lag values were evaluated: M ∈ {40, 72, 152, 252, 352, 600} ms, spanning cognitively-relevant timescales from early sensory processing (N1, 39–72 ms) to slow cognitive transitions (600 ms; 24). For node-level spatial analyses, row-sums of the 100 × 100 matrix were computed to generate n_"windows" ×100regional trajectories.

Trajectory descriptor extraction

From each TII trajectory, three scalar descriptors were derived per feature channel and averaged across the 28 network-pair connections:

**Temporal mean (mean_t):** average TII magnitude across time windows, indexing the mean hierarchical depth.

**Temporal variance (var_t):** variance across time windows, indexing the breadth of state-space exploration.

**Lag-1 autocorrelation (autocorr):** first-order temporal autocorrelation across sequential windows, indexing temporal persistence.

For cross-sectional analyses in the Isha cohort, eyes-closed segments were concatenated per subject. To prevent baseline-drift contamination of the second-order descriptors, time series were demeaned per segment prior to pooling before computing var_t and autocorr.

Statistical framework

All statistical tests used the Benjamini–Hochberg False Discovery Rate (BH-FDR) procedure for multiple-comparison control. The correction unit is stated for each analysis below. BH-FDR was preferred over Benjamini–Yekutieli because our dependent tests across lags and network pairs are expected to be positively correlated (all measures share the same underlying trajectory), which is precisely the regime where BH-FDR is valid and BY is unnecessarily conservative. Symbol convention throughout: * = BH-FDR-corrected significant ($q < 0.05$, with ** $q < 0.01$ and *** $q < 0.001$), and † = uncorrected significant only (raw $p < 0.05$, $q \geq 0.05$).

Cross-sectional group contrasts (Fig. 2a, Fig. 3a, Supplementary Fig. 1a). Differences between ADV and CNT in ISHA were evaluated using Welch's two-sample t-tests with Cohen's d effect sizes. Novice practitioners (NOV) are plotted as an intermediate visual reference between the two extreme groups; the primary contrast is ADV versus CNT (Welch's t-test) to avoid inflating multiplicity, but the NOV group falls consistently at intermediate values, supporting a continuous expertise-related gradient. Parametric Welch's tests were used for cross-sectional group contrasts, and non-parametric paired sign-flip permutation tests were used for within-subject longitudinal contrasts to accommodate the smaller CoMET sample (n = 17) without normality assumptions..

FDR correction was applied across the 6 lags (Fig. 2a) or across the 28 network pairs per (metric × lag) combination (Fig. 3a, Supplementary Fig. 1a).
Longitudinal within-subject contrasts (Fig. 2b, Fig. 3b–c, Supplementary Fig. 1b–c). Post − pre changes in the MBSR and CoMET cohorts were tested with paired sign-flip permutation (10,000 iterations), reporting the paired t-statistic and Cohen's $d_z$. FDR correction was applied across the 6 lags per cohort (Fig. 2b) or across the 28 network pairs per (metric × lag × cohort) combination (Fig. 3b–c, Supplementary Fig. 1b–c).

Cross-sectional dose–response (Fig. 5A). Spearman ρ between $\log_{10}(1+\text{"Practice Hours"})$and var_t at each of the six lags in the ISHA cohort (n = 96). Practice hours were assigned from the ISHA Master Sheet (0 to 15,400 h). BH-FDR correction across the 6 lags. Fisher-transform 95 % confidence intervals are shown on the ρ curve.

HMM state × trait descriptor correlations (Fig. 4B). Spearman ρ between per-subject state Fractional Occupancy (FO of each of the 5 states) and each of the three trait descriptors at each of the six TII lags, computed on the ISHA cross-sectional cohort only (n = 96, one observation per subject). ISHA-only was chosen to avoid the non-independence introduced by paired pre/post observations in the longitudinal cohorts. BH-FDR correction was applied globally across the 90 tests (5 states × 6 lags × 3 descriptors).

HMM longitudinal intervention effects on FO (Fig. 4C). Paired sign-flip permutation (10,000 iterations) on post − pre FO within each cohort (MBSR n = 28; CoMET, n = 17), with Cohen's $d_z$. BH-FDR correction across the 10 pre/post contrasts (5 states × 2 cohorts).

Longitudinal Δ Brain × Δ FFMQ (Fig. 5 rows B and C). For each subject in the pooled MBSR sample (n = 43; MBSR + CoMET, n = 42), we computed Δ trait descriptor (post − pre) and Δ FFMQ facet (post − pre), and Spearman ρ between the two at each lag per FFMQ facet. BH-FDR correction across the 6 lags per (facet × brain descriptor) combination.

<u>Hidden Markov Model training and state inference</u>
To resolve the microstate architecture underlying macroscopic trajectories, a joint Hidden Markov Model was trained on TII trajectories at the primary timescale (M = 152 ms) across all cohorts. The training set comprised 186 subject × condition segments (ISHA all-EC + MBSR pre + MBSR post + CoMET pre + CoMET post; n = 141 unique subjects). Segments shorter than 60 windows were excluded. Time series were per-segment demeaned prior to training. Models were trained using the osl_dynamics Python package (v1.2.x) with parameters: sequence_length = 50, batch_size = 32, learning_rate = $10^{-3}$, and a covariance-only emission model (learn_means = False, learn_covariances = True), chosen to isolate dynamic dispersion from mean-activity contributions. State dimensionality $K \in \{3, 4, 5, 6, 7\}$ was evaluated across 5 random initialisations (15 epochs each). K = 5 was selected on the basis of free-energy minimisation and Hamming-distance consistency . The final model was selected from the lowest-free-energy run among 30 random initialisations at K = 5 (Free Energy = 757.83). Viterbi decoding was applied to infer per-subject state sequences. The primary state-level parameter used throughout Fig. 4 is the Fractional Occupancy ($FO_k$), the proportion of total time each subject spent in state k. Complementary state-level quantities (Mean Dwell Time, Switching Rate, Transition Matrix Entropy) were computed for internal validation but are not shown in the main figures.

<u>Sensitivity and robustness analyses</u>
Robustness of the var_t effect was verified by comparing per-condition and pooled segment concatenation strategies in ISHA, yielding qualitatively identical directional effects (Supplementary Methods). Robustness of the autocorr effect to sliding-window overlap was verified by comparison with overlap-independent lag-10 autocorrelation on unbroken time series (Supplementary Fig. 2). Truncation of the MBSR and CoMET recordings to shorter durations preserved the direction of every reported effect (Supplementary Methods). The mean_t triple-negative (no expertise gradient in ISHA, no MBSR effect in the MBSR cohort, no CoMET effect) was confirmed at the connectome level across all 28 network pairs and all six lags (Supplementary Fig. 1, three-metric version of Fig. 3).

Data availability

De-identified subject-level metrics, extracted HMM parameters, and summary statistical tables supporting all figures are available in the online Source Data files accompanying this publication and archived on Zenodo (DOI: 10.5281/zenodo.21419256). Raw EEG datasets are subject to participant privacy agreements. Data-access requests may be directed to D.Z. (healthy MBSR cohort, Université Libre de Bruxelles), A.S.K. / H.F. (CoMET trauma-exposed cohort, Murdoch University), or the Sadhguru Center for a Conscious Planet (Isha Yoga cohort).

Code availability

All processing scripts, sliding-window TII computation, descriptor extraction, statistical testing, HMM training pipelines, and figure generation, are open-source under the MIT License on GitHub (https://github.com/LIKACT/tii_meditation) and archived on Zenodo (DOI: 10.5281/zenodo.21419256). The analysis package is structured as a modular Python library (tii_meditation, v1.0.0) with documentation and step-by-step reproduction instructions.

## Acknowledgments

We gratefully acknowledge all participants for their time, engagement, and commitment to this research. We thank the research teams in Brussels and Australia for generously sharing their datasets and for their contributions to data collection and study design.

**Figures and Tables**

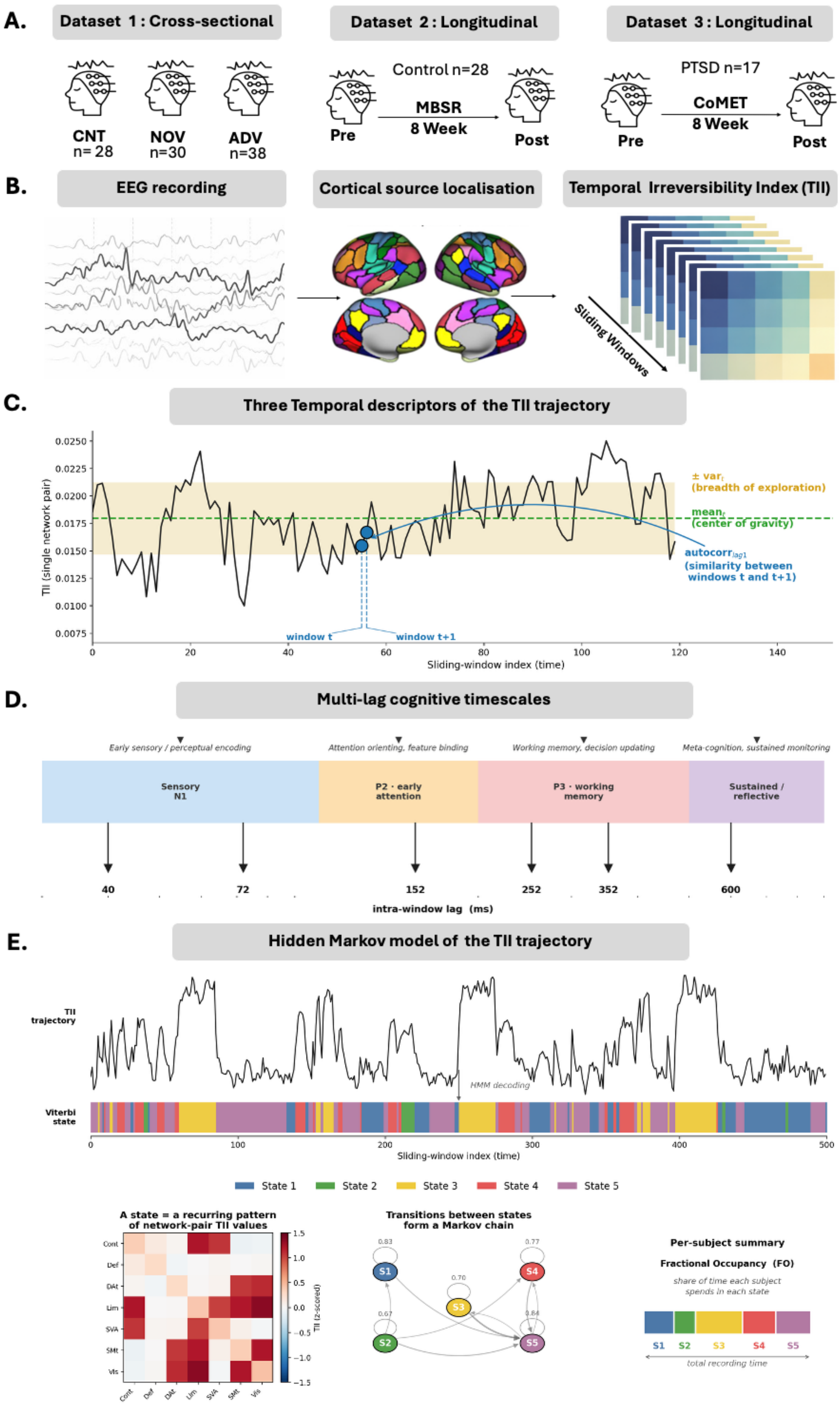
A.
Dataset 1 : Cross-sectional
CNT n= 28
NOV n=30
ADV n=38
Dataset 2 : Longitudinal
Control n=28
MBSR
8 Week
Pre
Post
Dataset 3 : Longitudinal
PTSD n=17
CoMET
8 Week
Pre
Post
B.
EEG recording
Cortical source localisation
Temporal Irreversibility Index (TII)
Sliding Windows
C.
Three Temporal descriptors of the TII trajectory
TII (single network pair)
± var
(breadth of exploration)
mean
(center of gravity)
autocorr
(similarity between windows t and t+1)
window t
window t+1
Sliding-window index (time)
D.
Multi-lag cognitive timescales
Early sensory / perceptual encoding
Attention orienting, feature binding
Working memory, decision updating
Meta-cognition, sustained monitoring
Sensory N1
P2 · early attention
P3 · working memory
Sustained / reflective
40
72
152
252
352
600
intra-window lag (ms)
E.
Hidden Markov model of the TII trajectory
TII trajectory
HMM decoding
Viterbi state
Sliding-window index (time)
State 1
State 2
State 3
State 4
State 5
A state = a recurring pattern of network-pair TII values
Cont
Def
DAt
Lim
SVA
SMt
Vis
TII (z-scored)
Transitions between states form a Markov chain
S1
S2
S3
S4
S5
Per-subject summary
Fractional Occupancy (FO)
share of time each subject spends in each state
total recording time

**Figure 1.** Study design, sliding-window TII, and analytical framework. (a) Multi-cohort dataset (n=141). Three independent EEG cohorts were evaluated under a unified pipeline: a cross-sectional Isha Yoga cohort (n=96; non-meditator controls [CNT], novices [NOV], and advanced practitioners [ADV]), a longitudinal healthy MBSR cohort (n=28; pre- vs. post-8-week intervention), and a longitudinal trauma-exposed cohort undergoing CoMET; n=17; pre- vs. post-8-week intervention). (b) Computational pipeline. Sensor EEG signals were source-reconstructed onto the 100-region Schaefer cortical parcellation grouped into the seven canonical Yeo functional networks. For each participant and maximum lag M, bivariate Temporal Irreversibility Index (TII) matrices were computed within sliding time windows (length 5M, step M/2) and condensed into 27-dimensional inter-network feature vectors over time. (c) Trajectory descriptors. Each TII trajectory was parameterized by three scalar metrics characterizing its dynamic regime: temporal mean (hierarchical depth; average hierarchical depth), temporal variance (exploratory breadth; state-space exploration breadth), and lag-1 temporal persistence (autocorr; temporal persistence across consecutive windows). (d) Evaluated timescales. Six intra-window maximum lags M were probed, ranging from 40 to 600 ms, capturing distinct neurocognitive processes, including early sensory processing (N1, 39–72 ms), attentional selection (P2, 152 ms), working memory updating (P3, 252–352 ms), and sustained metacognitive monitoring (600 ms). (e) Joint Hidden Markov Model (HMM). A 5-state Gaussian HMM was trained on TII trajectories at the primary timescale (M=152 ms) across all three cohorts. State sequences decoded via the Viterbi algorithm (top) reveal recurring 27-dimensional network-pair covariance patterns (bottom left) governed by a sparse Markov transition matrix (bottom middle), yielding per-participant Fractional Occupancy ("FO" ) estimates across states (bottom right).

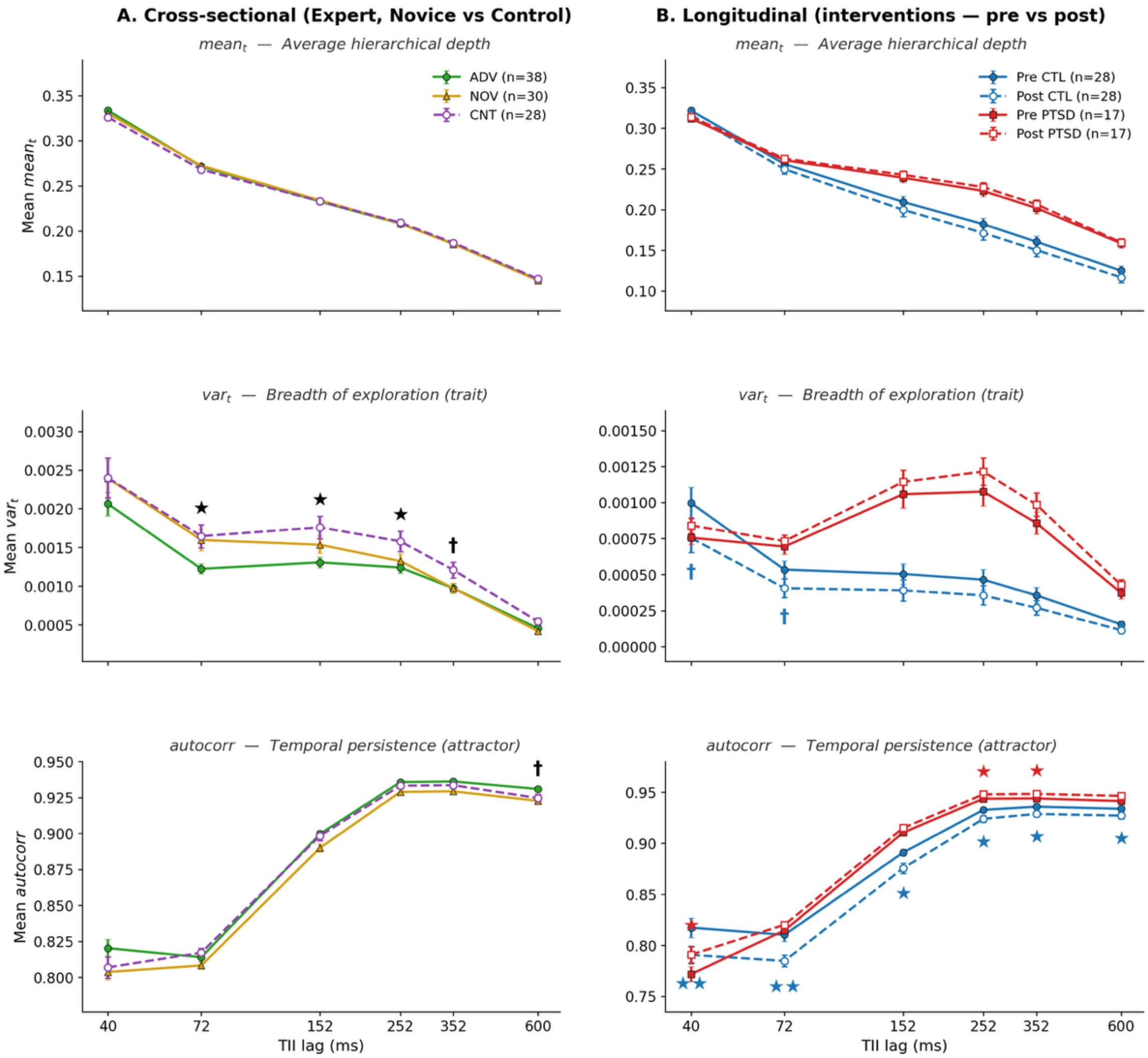

A. Cross-sectional (Expert, Novice vs Control)
B. Longitudinal (interventions — pre vs post)
$mean_t$ — Average hierarchical depth
ADV (n=38)
NOV (n=30)
CNT (n=28)
Pre CTL (n=28)
Post CTL (n=28)
Pre PTSD (n=17)
Post PTSD (n=17)
Mean $mean_t$
$var_t$ — Breadth of exploration (trait)
Mean $var_t$
autocorr — Temporal persistence (attractor)
Mean autocorr
TII lag (ms)
40
72
152
252
352
600

**Figure 2.** Three temporal descriptors of hierarchical brain dynamics dissociate across experimental designs. Matrix layout: Rows display the three trajectory descriptors (hierarchical depth, top; exploratory breadth, middle; temporal persistence, bottom); columns represent experimental designs. Data are presented as mean ± s.e.m. across intra-window TII lags $\tau \in \{40,72,152,252,352,600\}$ ms. (a) Cross-sectional expertise gradient (Isha Yoga cohort). Advanced practitioners (ADV, n=38; dark green), novices (NOV, n=30; gold), and non-meditator controls (CNT, n=28; purple dashed). Metrics were computed on baseline resting-state EEG (pooled PreMed + PreSitStill eyes-closed segments, ~4 min per subject). Statistical comparison: Welch's two-sample t-test (ADV vs. CNT) at each lag with Benjamini–Hochberg False Discovery Rate correction across the six lags (BH-FDR, q<0.05). While global hierarchical depth (hierarchical depth) displays no expertise effect, exploratory breadth (exploratory breadth) shows a robust expertise-dependent elevation (⋆ at 72, 152, and 252 ms; † at 352 ms). temporal persistence shows only a marginal trait effect at long timescales († at 600 ms). (b) Longitudinal intervention dynamics (MBSR healthy vs. CoMET trauma-exposed cohorts). Healthy MBSR meditators (n=28; blue circles, solid pre- vs. open dashed post-intervention) and trauma-exposed individuals with PTSS undergoing 8-week CoMET therapy (n=17; red squares, solid pre- vs. open dashed post-intervention). Statistical comparison: paired sign-flip permutation tests (10,000 iterations) on post - pre differences per subject, BH-FDR corrected across the six lags within each cohort. Significance markers are color-coded by cohort (blue: healthy MBSR; red: CoMET (trauma-exposed)). Key findings: hierarchical depth is invariant to both mindfulness interventions. exploratory breadth selective reduction: healthy meditators exhibit a decrease at fast lags († at 40 and 72 ms), whereas no change occurs in the trauma-exposed cohort, reflecting a healthy-specific reduction in state-space exploratory breadth. temporal persistence bidirectional reorganization: healthy MBSR participants show a pan-lag reduction in temporal persistence (⋆⋆ at 40 and 72 ms; ⋆ at 152, 252, 352, and 600 ms), whereas trauma-exposed individuals with PTSS display an increase post-intervention (⋆ at 252 and 352 ms). Significance convention (applies to all figures): All P-values are corrected for multiple comparisons via BH-FDR at the level specified in each legend. Symbols denote evidence levels: ⋆ FDR-corrected significance (q<0.05; ⋆⋆ q<0.01; ⋆⋆⋆q<0.001), and † uncorrected significance (P<0.05, q≥0.05).

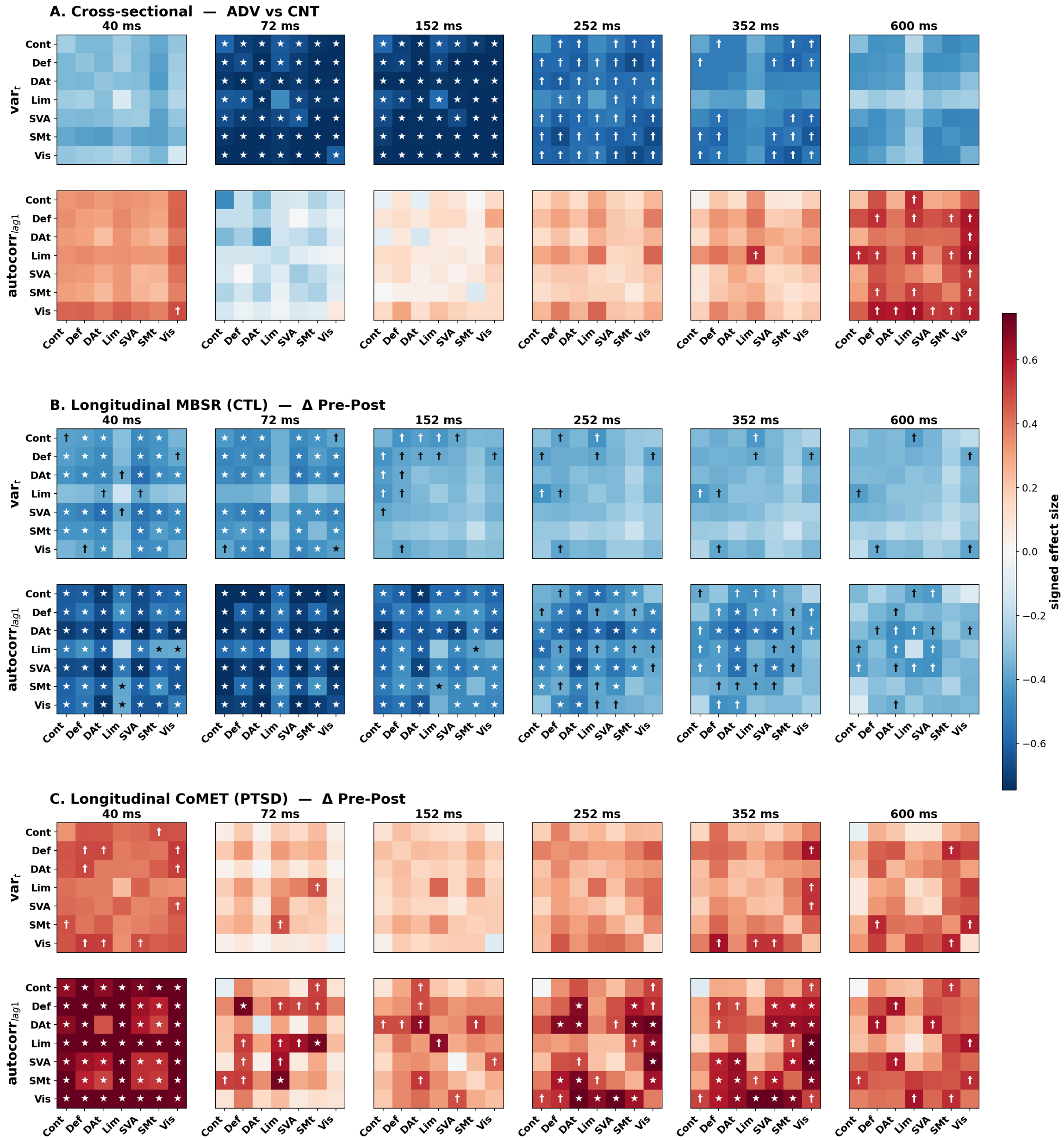

A. Cross-sectional — ADV vs CNT
B. Longitudinal MBSR (CTL) — Δ Pre-Post
C. Longitudinal CoMET (PTSD) — Δ Pre-Post
40 ms
72 ms
152 ms
252 ms
352 ms
600 ms
var_t
autocorr_lag1
Cont
Def
DAt
Lim
SVA
SMt
Vis
signed effect size
0.6
0.4
0.2
0.0
−0.2
−0.4
−0.6

**Figure 3.** Spatial distribution of network effects: pan-cerebral modulation in healthy meditators versus focal reorganization in trauma-exposed patients. Connectome layout: Three major sections organize the spatial effect maps by cohort: (A) cross-sectional Isha Yoga (ADV vs. CNTΔ), (B) longitudinal healthy MBSR (MBSR cohort; post − pred_z), and (C) longitudinal CoMET (trauma-exposed cohort; post − pred_z). Within each cohort section, rows display two trajectory descriptors (exploratory breadth, top; temporal persistence, bottom) across six columns corresponding to the probed TII intra-window lags (39, 72, 152, 252, 352, and 600 ms). hierarchical depth is omitted due to a lack of significant spatial modulation across all cohorts (all q>0.10; see Supplementary Fig. 1). Each matrix cell displays a 7×7 heatmap of inter-network effect sizes among the seven canonical Yeo functional networks (Control [Cont], Default Mode [Def], Dorsal Attention [DAt], Limbic [Lim], Salience/Ventral Attention [SVA], Somatomotor [SMt], and Visual [Vis]), yielding 28 unique inter-network pairs symmetrically mirrored across the diagonal. Color scale indicates effect direction: blue = decrease post-intervention or lower in advanced practitioners (ADV < CNT); red = increase post-intervention or higher in advanced practitioners (ADV > CNT). Statistical significance was controlled using Benjamini–Hochberg False Discovery Rate correction applied across the 28 network pairs per condition (metric × lag × cohort): ⋆ FDR-corrected significance (q<0.05); † uncorrected significance (P<0.05, q≥0.05). Key findings: Pan-cerebral vs. localized exploratory breadth modulation: In the Isha cohort, advanced meditators exhibit widespread, pan-cerebral reductions in exploratory breadth relative to controls (ADV < CNT) peaking at intermediate lags (72 and 152 ms; virtually all 28 pairs reaching q<0.05). Similarly, healthy MBSR participants (CTL) display pan-cerebral post-intervention decreases at fast lags (40 and 72 ms). Conversely, trauma-exposed individuals with PTSS demonstrate a directional reversal, showing elevated exploratory breadth tendencies across network pairs († uncorrected at 40 ms). Pan-cerebral vs. focal temporal persistence reorganization: Healthy MBSR participants (CTL) demonstrate a global, pan-cerebral reduction in temporal persistence (26 of 28 network pairs reaching q<0.05 across all six timescales). In contrast, post-CoMET trauma-exposed individuals with PTSS display a spatially localized increase in temporal persistence, concentrated within Dorsal Attention × Sensorimotor, Control × Default Mode, and intra-Limbic network couplings († uncorrected at 252–352 ms).

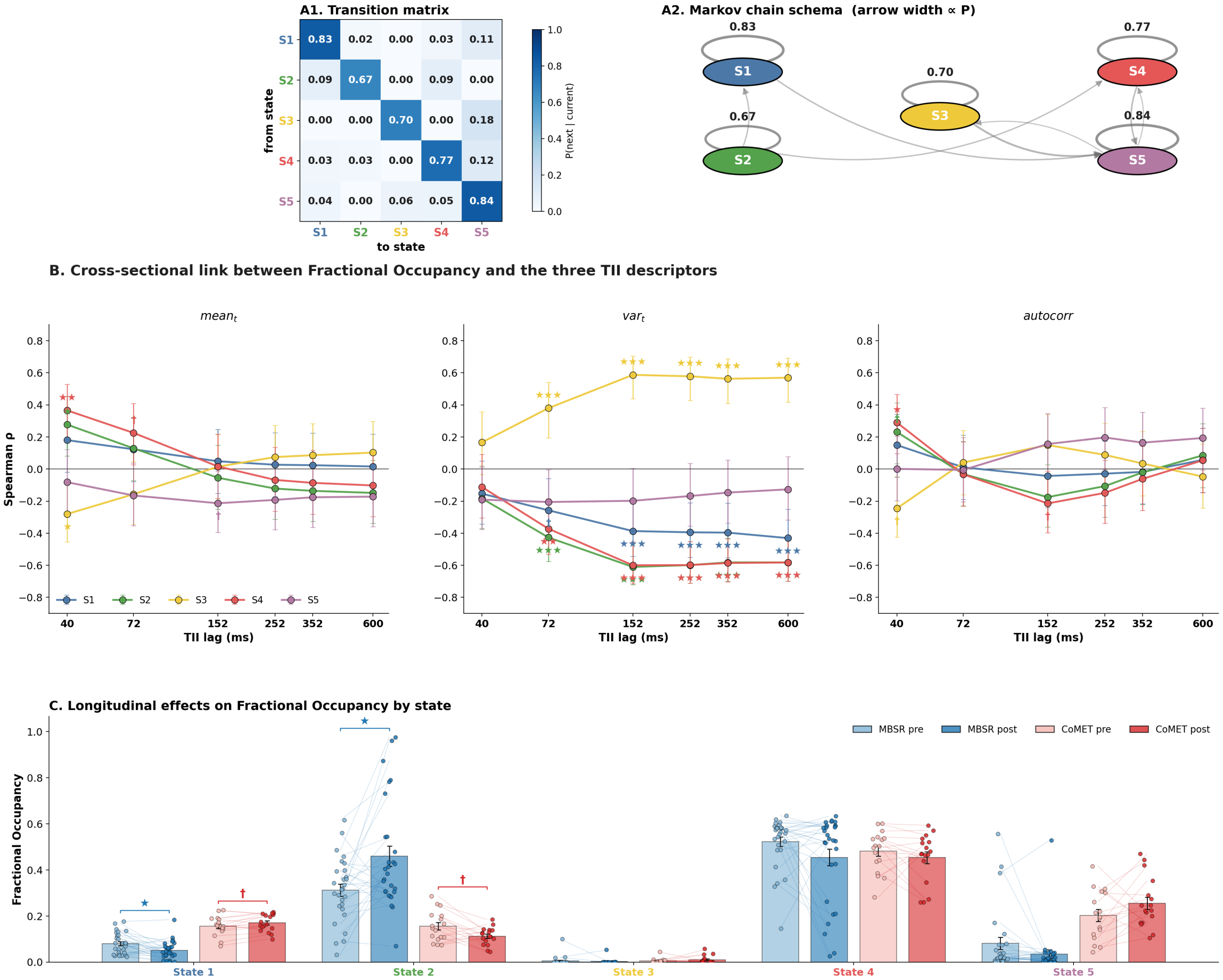
A1. Transition matrix
A2. Markov chain schema (arrow width ∝ P)
B. Cross-sectional link between Fractional Occupancy and the three TII descriptors
C. Longitudinal effects on Fractional Occupancy by state

**Figure 4.** Hidden Markov Model reveals that Fractional Occupancy tracks the exploratory-breadth axis at the trait level, and a state-level dissociation between healthy and trauma-exposed populations after intervention. (A) Joint 5-state HMM architecture. A joint Gaussian HMM was trained on TII trajectories at the primary timescale (M=152 ms) across all three cohorts (n=186 total subject × condition segments; ISHA all-EC + MBSR pre/post + CoMET pre/post). (A1) Transition probability matrix displaying5×5 high self-transition probabilities ("sticky" dynamics: diagonal P=0.83,0.67,0.70,0.77,0.84 for States 1–5, respectively) and sparse inter-state transitions. (A2) Markov chain graph representation, where node colors match state designations throughout the figure and edge widths are proportional to transition probabilities. (B) Cross-sectional link between Fractional Occupancy (FO) and trajectory descriptors (Isha Yoga cohort, n=96). Spearman rank correlations betweenρ per-subject state "FO"  and each trait descriptor across the six probed TIIlags (40 to 600 ms). To prevent non-independence issues, only cross-sectional Isha Yoga participants were included (n=96, single observation per subject). Error bars denote 95% confidence intervals estimated via Fisher z-transformation. Multiple comparisons were controlled globally using Benjamini–Hochberg False Discovery Rate correction across all 90 tests (5 states × 6 lags × 3 axes; q<0.05). Exploratory breadth is the sole axis coupling robustly to state occupancy: State 3 exhibits a large, sustained positive contribution across 152–600 ms (ρ≈+0.60, all ⋆⋆⋆), whereas States 2 and 4 contribute equally strong negative signals over the same timescales (ρ≈-0.60, all ⋆⋆⋆), alongside a moderate negative contribution from State 1 (ρ up to -0.42 at 600 ms, ⋆⋆). In contrast, hierarchical depth and temporal persistence remain tightly bounded near zero across nearly all states and lags (only State 4 at 40 ms reaches significance: ⋆⋆ and ⋆, respectively). This selective dissociation establishes  as a state-occupancy phenomenon, where participants spending more time in the sparse integrative State 3 and less time in persistent anchor States 2 and 4 display systematically higher trajectory variance, whereas  and  reflect complementary dynamic features (such as dwell-time distributions and transition rhythms) uncaptured by static occupancy. (C) Longitudinal state-level intervention dynamics. Pre- (light bars) and post-intervention (dark bars) Fractional Occupancy for"FO"  healthy MBSR (n = 28, n=28; blue) and CoMET cohorts (n=17; red). Dots represent individual participants, thin lines connect paired pre/post observations, and error bars denote s.e.m. Statistical testing: paired sign-flip permutation tests (10,000 iterations) with BH-FDR correction applied across the 10 pre/post contrasts (5 states × 2 cohorts). Key finding: Healthy and clinical populations demonstrate opposing state-level reorganizations within the two low-occupancy states: State 1: Decreases in healthy meditators post-MBSR (⋆, q<0.05), but increases in trauma-exposed individuals with PTSS post-CoMET (†uncorrected, P<0.05). State 2: Increases in healthy meditators post-MBSR (⋆, q<0.05), but decreases in trauma-exposed individuals with PTSS post-CoMET (†uncorrected, P<0.05). States 3, 4, and 5 exhibit no significant intervention-related shifts in either cohort. This state-level divergence in States 1 and 2 directly underpins the population-specific responses observed for (Fig. 2b) and inter-network coupling profiles (Fig. 3).

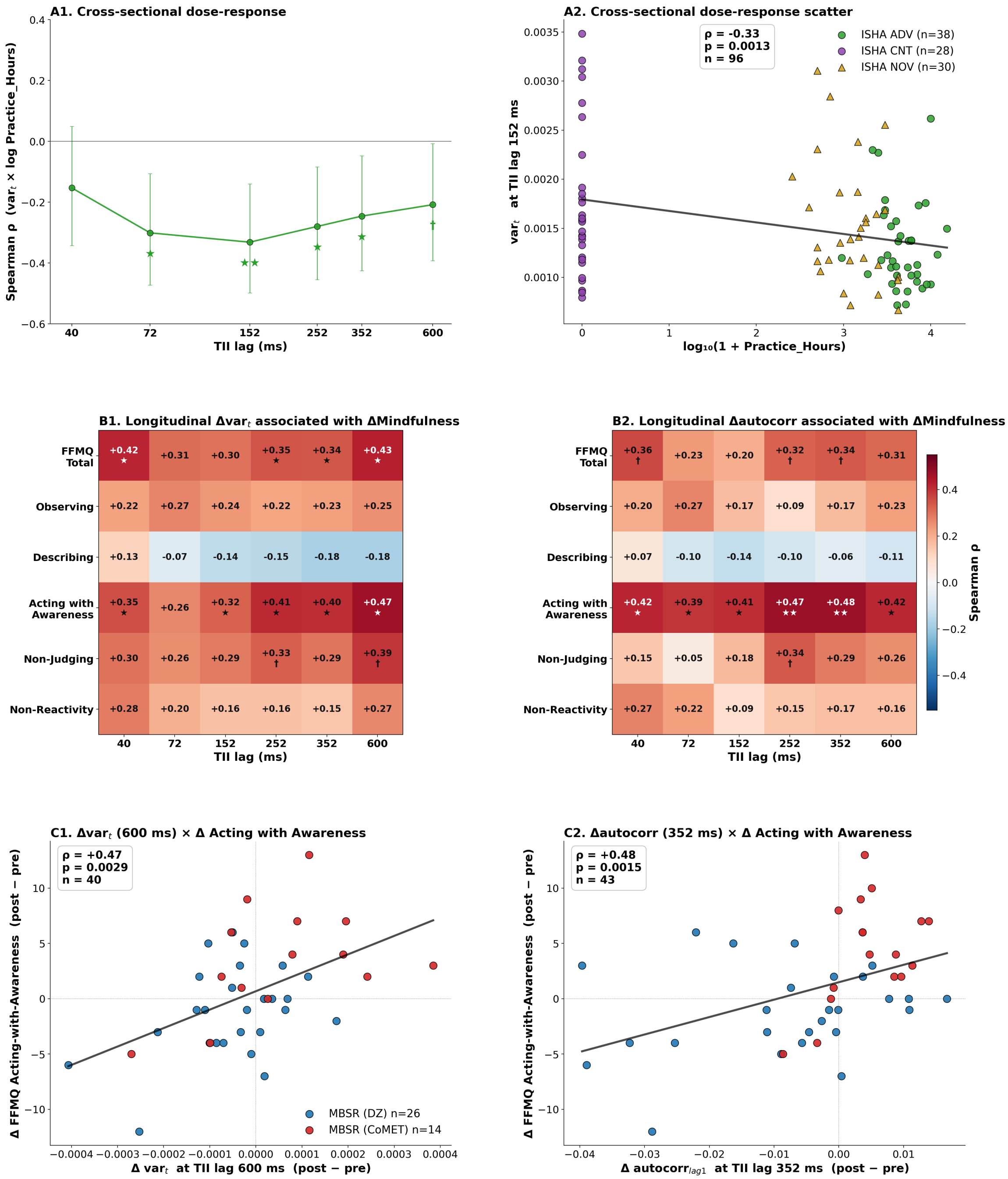
A1. Cross-sectional dose-response
Spearman ρ (var$_t$ × log Practice_Hours)
TII lag (ms)
A2. Cross-sectional dose-response scatter
ρ = -0.33
p = 0.0013
n = 96
ISHA ADV (n=38)
ISHA CNT (n=28)
ISHA NOV (n=30)
var$_t$ at TII lag 152 ms
log₁₀(1 + Practice_Hours)
B1. Longitudinal Δvar$_t$ associated with ΔMindfulness
B2. Longitudinal Δautocorr associated with ΔMindfulness
FFMQ Total
Observing
Describing
Acting with Awareness
Non-Judging
Non-Reactivity
TII lag (ms)
Spearman ρ
C1. Δvar$_t$ (600 ms) × Δ Acting with Awareness
ρ = +0.47
p = 0.0029
n = 40
MBSR (DZ) n=26
MBSR (CoMET) n=14
Δ FFMQ Acting-with-Awareness (post − pre)
Δ var$_t$ at TII lag 600 ms (post − pre)
C2. Δautocorr (352 ms) × Δ Acting with Awareness
ρ = +0.48
p = 0.0015
n = 43
Δ autocorr$_{lag1}$ at TII lag 352 ms (post − pre)

**Figure 5.** Behavioral bridges: cumulative practice dose and Acting-with-Awareness track state-space exploration and temporal persistence. Row A. Cross-sectional dose–response relationship (Isha Yoga cohort, n=96). (A1) Multi-scale dose–response profile. Spearman rank correlation between$\rho$ log-transformed cumulative practice hours and$\log_{10}[1+\text{"hours"}]$ exploratory breadth across the six TII intra-window lags. Shaded envelope indicates 95% confidence intervals estimated via Fisher z-transformation. Significance was controlled using Benjamini–Hochberg False Discovery Rate correction across the six lags (BH-FDR, $q<0.05$). displays a robust, negative dose–response relationship spanning four consecutive timescales (72–352 ms, all $q<0.05$), peaking at M=152 ms ($\rho$=-0.33, ⋆⋆). (A2) Dose–response scatter plot at the peak timescale (M=152 ms). Individual participants are color-coded by expertise level: non-meditator controls (CNT, purple), novices (NOV, gold), and advanced practitioners (ADV, green). The steady negative gradient across four orders of magnitude of practice (0 to 15,400 h; global $\rho$=-0.33, P=0.0013) supports a continuous, dose-dependent trait modulation of state-space exploration. Row B. Longitudinal brain–psychometric coupling (Δ Brain × Δ FFMQ , pooled MBSR sample, n=43). Within-subject changes were$\Delta=\text{"post"}-\text{"pre"}$ computed for neural trajectory metrics and Five-Facet Mindfulness Questionnaire (FFMQ) subscales in the combined MBSR dataset (healthy MBSR, n=28; CoMET (trauma-exposed), n=15 with complete psychometrics). Heatmaps display Spearman rank correlations across six TII timescales for total score and the five FFMQ facets (Observing, Describing, Acting-with-Awareness, Non-Judging, Non-Reactivity), with BH-FDR correction applied across the six lags per facet. (B1) : The Acting-with-Awareness subscale concentrates the association, reaching significance across five of six lags (peaking at M=600 ms: $\rho$=+0.47, ⋆). FFMQ Total score also shows significant alignment across four lags. (B2) : Demonstrates stronger, pan-lag alignment specifically within the Acting-with-Awareness dimension, reaching FDR significance across all six timescales and peaking at intermediate-to-long lags (M=252"--" 352 ms: $\rho$=+0.47 and +0.48, ⋆⋆). No other FFMQ facet reaches multi-lag FDR significance. Row C. Scatter plots of primary brain–behavior associations (pooled MBSR sample). (C1) at M=600 ms versus Δ Acting-with-Awareness ($\rho$=+0.47, P=0.0029, n=40; three CoMET participants had insufficient trajectory length for reliable exploratory breadth computation at the longest lag (M = 600 ms) and were excluded). (C2) at M=352 ms versus Δ Acting-with-Awareness ($\rho$=+0.48, P=0.0015, n=43). Individual participants are color-coded by cohort (healthy MBSR, blue; CoMET (trauma-exposed), red). Both healthy and clinical participants adhere to the same underlying regression slope, indicating that this behavioral anchor operates independently of clinical baseline status.